\documentclass[%
reprint,
amsmath,amssymb,
aps,
prd,
showpacs,
floatfix,
]{revtex4-2}

\usepackage{graphicx}
\usepackage{dcolumn}
\usepackage{bm}
\usepackage[colorlinks=true,citecolor=blue,linkcolor=blue,urlcolor=blue]{hyperref}

\usepackage{tikz}
\usepackage{xcolor}
\usepackage{soul}

\newcommand{\orcidicon}[1]{\href{https://orcid.org/#1}{\includegraphics[height=\fontcharht\font`\B]{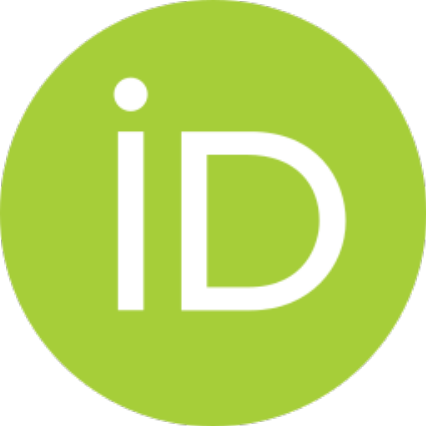}}}

\usepackage{mathrsfs}
\usepackage{pstricks}
\usepackage{color}
\usepackage{slashed}
\usepackage{epsfig}
\usepackage{amsfonts}
\def\beq{\begin{equation}}
\def\eeq{\end{equation}}
\def\bea{\begin{eqnarray}}
\def\eea{\end{eqnarray}}
\def\nn{\nonumber}

\def\Re{\textrm{Re}}

\def\d{\mathrm{d}}
\def\q_perp{\mathbf{q}_{\perp}}
\begin{document}

\title{Critical Casimir effect in a disordered $O(2)$-symmetric model}

\author{G. O. Heymans\,\orcidicon{0000-0002-1650-4903 }}
\email{Email address: olegario@cbpf.br}
\affiliation{Centro Brasileiro de Pesquisas F\'{\i}sicas - CBPF, \\ Rua Dr. Xavier Sigaud 150 22290-180 Rio de Janeiro, RJ, Brazil
}
\author{N.~F.~Svaiter\,\orcidicon{0000-0001-8830-6925}}

\email{Email address: nfuxsvai@cbpf.br}
\affiliation{Centro Brasileiro de Pesquisas F\'{\i}sicas - CBPF, \\ Rua Dr. Xavier Sigaud 150 22290-180 Rio de Janeiro, RJ, Brazil
}

\author{B.~F.~Svaiter\,\orcidicon{0000-0003-3656-9883}}
\email{Email address: benar@impa.br}
\affiliation{Instituto de Matemática Pura e Aplicada - IMPA \\ Estrada Dona Castorina 110 22460-320 Rio de Janeiro, RJ, Brazil}

\author{G.~Krein\,\orcidicon{0000-0003-1713-8578}}
\email{Email address: gastao.krein@unesp.br}
\affiliation{Instituto de F\'{i}sica Te\'orica, Universidade Estadual Paulista - IFT-UNESP\\
Rua Dr. Bento Teobaldo Ferraz, 271, Bloco II 01140-070 S\~ao Paulo, SP, Brazil}
%


\begin{abstract}
Critical Casimir effect appears when critical fluctuations
of an order parameter interact with classical boundaries. 
We investigate this effect in the setting of a Landau-Ginzburg 
model with continuous symmetry in the presence of quenched disorder.
The quenched free energy is written as an asymptotic series of moments of 
the model's partition function. 
Our main result is that, in the presence of a strong disorder, 
Goldstone modes of the system contribute either with an attractive 
or with a repulsive force.
This result was obtained using the 
distributional zeta-function method without relying  on any 
particular ansatz in the functional space of the moments of 
the partition function.  
\end{abstract}


\maketitle



\section{Introduction}\label{intro}

Quantum fields are mathematical objects that allow a general 
description of the physical world.
%
%
In the axiomatic approach they are operator-valued generalized functions 
acting over test function 
spaces \cite{gelfand,Streater:1989vi}.
Such a description leds the local energy of 
quantum fields to attain negative values~\cite{Epstein:1965zza}.
In the presence of boundaries, negative local energies generate 
attractive forces.
This result, known in the literature as the Casimir 
effect~\cite{Casimir:1948dh, ca1}, manifests itself for all types of 
fundamental fields, scalar, fermionic, and vector 
fields~\cite{Plunien:1986ca,mostepanenko1997casimir,Milton2001,Fermi:2015xua}.


In an Euclidean functional integral description, due the randomness 
properties of quantum fields, they need to be integrated over the functional 
space \cite{Gelfand:1959nq}. From such a functional/classical probabilistic 
point of view, it is known that if the mean of a nonzero random variable 
vanishes, their variance differs from zero. This fact alone suffices to give rise to Casimir forces. The physical
reason behind the Casimir effect can be traced 
to the presence of massless excitations
and the change of the thermodynamic equilibrium of the vacuum (state 
with zero number occupation) due to the presence of boundaries that 
change the fluctuating spectrum of the 
theory \cite{Fulling:1989nb}.

Holding the physical interpretation of the Casimir forces, one can expect that a similar effect happens for critical systems with infinite correlation 
lengths in the presence of boundaries. Such a situation was discussed in fluids first by Fisher and de Gennes~\cite{fisher2003phenomenes}. As a matter of fact, thermal fluctuations can induce Casimir-like long-ranged forces in any correlated medium, with a critical system being an example. In such a situation the massless excitations are not associated with photons but with some other quasi-particles, \textit{e.g.}, phonons or Goldstone bosons. Such an effect is referred to as the critical or the thermodynamic Casimir effect. So far the critical Casimir effect has enjoyed only a few reviews, \textit{e.g.} those of Refs.~\cite{Krech:1994, brankov2000theory, gambassi2009casimir,dantchev2023critical, gambassi2023critical}.

%
Quantum Nyquist theorem~\cite{ginzburg1987quantum} allows one to identify 
regimes where thermal fluctuations dominate over those of quantum origin, 
with the possibility of systems becoming critical. Such situations are the 
subject of 
statistical field theory. When a system reaches the critical regime, correlations 
become long-ranged, and critical Casimir  forces appear. Besides thermal 
fluctuations, disorder fluctuations can also drive a system to 
criticality~\cite{nattermann1989random}. A prototype model featuring disorder 
fluctuations is a binary fluid in a porous medium~\cite{broch}, whose critical 
behavior can be studied as a continuous field in the presence of a random field. 
When the binary-fluid correlation length is smaller than the porous radius, one 
has a system with finite-size effects in the presence of a surface field. When the 
binary fluid correlation length is much larger than that of the porous radius, the 
random porous can exert a random field effect. In the latter case, introduction of 
boundaries gives rise to the critical Casimir effect~\cite{gambassi2009casimir}.

A similar scenario, but with a discrete symmetry, was studied in 
Ref. \cite{Rodriguez-Camargo:2022wyz}.
The main result of that study was that a change 
in the sign ofthe Casimir force
can happen depending on the ratio of the inverse of the 
correlation length and the disorder strength. This result is analogous 
to the situation of the electromagnetic Casimir effect which
can change sign depending on the ratio between the permeability and 
the dielectric constant~\cite{boyer1974van}, disorder fluctuations 
lead to a Casimir force that is attractive or repulsive. We would like to point that there are some papers in the literature 
discussing the critical Casimir effect with the disorder at the surface, see e.g. \cite{maciolek2015critical, escidoc:2539714}.


The purpose of this work is to revisit the Casimir effect in disordered systems with a continuous symmetry.  More specifically, we consider continuous fields that model order parameters possessing a continuous symmetry in scenarios where the disorder fluctuations dominate over the thermal fluctuations. Examples of systems realizing such a scenario include a binary fluid in the presence of an external random field in the critical regime, superfluids, and liquid crystals. In such a situation, when the criticality is reached, one has to take into account the soft modes (Goldstone bosons) due to the symmetry breaking \cite{li1991fluctuation,li1992fluctuation}. 
Another difference, and perhaps a more technical one, is that in the approach 
that we adopt here  we \textit{do not choose any ansatz for the functional space}
in the series of the quenched free energy. This procedure will be clarified along 
the following sections. Our primary aim in this work is to answer whether the 
soft modes associated with the Goldstone boson favor or suppress the Casimir 
force and whether they affect the sign of the force. The result that we obtain 
for such question is that the soft modes do not 
affect the change of the sign of the force. However 
a interesting effect due the disorder arises. In the regime of strong 
disorder, where we only have the Casimir effect due the presence of the 
soft mode, the Goldstone mode contribution may 
change from attractive to repulsive. 
In other words, the presence of disorder may change the sign of the ``universal amplitude" 
due to the Goldstone modes.

The paper is organized as follows. In the firsts two sections we introduce the two 
main mathematical tools utilized in the paper: Sec. \ref{sec:speczeta} presents 
the spectral zeta-function regularization method and how one can use it to obtain the Casimir energy of a system, while in 
Sec. \ref{sec:distzeta} we introduce the distributional zeta-function method to 
evaluate the quenched free energy, revisiting the critical Casimir force 
due the disorder. In the Sec. \ref{sec:main} we presents our main results 
and calculations. Sec. \ref{sec:conc} contains our main conclusions 
alongside with future perspectives.

\section{Casimir Energy And Spectral Zeta-Function Regularization}\label{sec:speczeta}

\vspace{0.5cm}

In quantum field theory, Casimir force can be computed by analyzing either the local energy density~\cite{Brown:1969na, Bender:1976wb, Milton:1978sf, Kay:1978zr, Hays:1979bc, Actor:1994wv, Rodrigues:2003du}, or the total energy~\cite{Fierz:1960zq, Boyer:1968uf, Boyer:1970bb} of the quantized fields.
In this section we study the Casimir energy of a statistical field theory model describing a Gaussian scalar field $\phi(x_1, \dots, x_d)$  in a slab geometry with one compactified dimension, $\Omega_L \equiv \mathbb{R}^{d-1}\times [0,L]$. For simplicity, we assume Dirichlet boundary conditions:
\begin{equation}
    \phi(x_1, \dots, x_{d-1}, 0) = \phi(x_1, \dots, x_{d-1}, L) = 0.
    \label{bc-phi}
\end{equation}
%
We start discussing the scalar field satisfying Dirichlet boundary conditions inside a box with sides $L_1$, $L_2$,\dots,$L_d$.
The partition function of the theory is
%
\begin{equation}
    Z = \int_{\Omega} [\d \phi] \; e^{-\frac{1}{2}\int \d^{d} x \,\, \phi(x)\left( - \Delta + m_0^2\right) \phi (x)},
\end{equation}
where $\Omega$ in the integral specifies the space of fields satisfying the boundary conditions, $[\d \phi] \equiv \prod_{x \in \Omega}\d\phi(x)$ is a formal measure over the space of functions $\Omega$, $-\Delta$ is the Laplace operator, and $m_0^2$ the bare mass of the free field.
Since the action is quadratic in the fields, 
the functional integral can be evaluated, yielding
%
%
\begin{equation}\label{eq:fundet}
    Z = \left[\mathrm{det}(-\Delta + m_0^2)_{\Omega}\right]^{-\frac{1}{2}},
\end{equation}
where we omitted a normalization factor due to the total volume of the functional space and the symbol~$\Omega$ indicates the boundary condition under which the determinant must be computed. Using the fact that a positive semi-definite self-adjoint operator satisfies an eigenvalue equation, we can write such a determinant as
\begin{equation}
\mathrm{det}(-\Delta + m_0^2)_{\Omega} = \prod_{i=1}^{\infty} \lambda_i,
\label{det-lambda}
\end{equation}
with the set of all $\lambda_i$ being the spectrum defined by the operator and its boundary condition. Equation~(\ref{det-lambda}) is formally divergent and requires regularization. We use the spectral zeta-function regularization method~\cite{Ray:1973sb, Hawking:1976ja,voros1992spectral, elizalde1995ten, Kirsten:2001wz}.
The zeta function regularization procedure is a special case of analytic regularization.
The use of the latter regularization in Casimir effect was discussed in Ref.~\cite{Ruggiero:1977eu, Ruggiero:1979ec}
References~\cite{Svaiter:1989gz,Svaiter:1991je,Svaiter:1993nk,Svaiter:1994vd} compare results for the Casimir energy obtained with an analytic regularization procedure and the traditional regularization using cutoff .  

To give a meaning to Eq.~(\ref{det-lambda}), one starts defining the spectral zeta-function, $\zeta_D(s)$, first for $\Re(s)>d/2$ as
\begin{equation}
  \label{eq:dsum}
    \zeta_D(s) \equiv \sum_{i=1}^{\infty}\frac{1}{\lambda_i^s},
\end{equation}
%
where $D$ specifies the differential operator under consideration. 
Second, extend it analytically to a maximal domain.
Observe that zero belongs to its domain.
Formally, from equation~\eqref{eq:dsum}, 
%
\begin{equation}
    \left.\frac{\d}{\d s}\zeta_D(s)\right|_{s=0} = -\sum_{i=1}^{\infty} \ln \lambda_i.
    \label{der-zeta}
\end{equation}
One can combine Eqs.~(\ref{det-lambda}) and (\ref{der-zeta}) to write the partition function in Eq.~(\ref{eq:fundet}) as
\begin{equation}
    Z = \exp\left(-\frac{1}{2}\sum^\infty_{i=1} \ln \lambda_i\right) = \exp\left(\frac{1}{2}\left.\frac{\d}{\d s}\zeta_D(s)\right|_{s=0}\right).
\end{equation}

To proceed with the calculations, we must construct the appropriate $\zeta_D(s)$. 
It can be constructed by using the appropriate spectral measure in the Riemann-
Stieljes integral. All the information about the domain $\Omega_L$ and the 
boundary conditions are  taken into account by the 
spectral measure. So, in the continuous limit, one obtains 
$\zeta_D(s)$ as:
%
\begin{equation}\label{eq: zetad}
    \zeta_D(s) = \frac{A_{d-1}}{(2\pi)^{d-1}}\int \d^{d-1}p \sum_{n=1}^{\infty} \left[p^2 + m_0^2 + \left(\frac{\pi n}{L}\right)^2 \right]^{-s},
\end{equation}
where $p^2 = p_1^2 + \cdots p_{d-1}^2$ and $A_{d-1}$ is the area of the hypersurface in $d-1$ dimensions:
\begin{equation}
    A_{d-1} \equiv \prod_{i=1}^{d-1}\lim_{L_i \to \infty} L_{i},
\end{equation}
where such a limit must be understood as $L_{i} \gg L, \,\, 
\forall \,\,i=1, \cdots, d-1$. From here on, one could proceed with the exact 
calculations of Ref. \cite{Hawking:1976ja}; see also Ref. \cite{Dittrich:1985yb}. 
In the following we introduce the calculation method we will use later on 
Sec.~\ref{sec:main}.
%
Such a method will reproduce the result in the literature via direct calculations. To proceed, let us to use that
\begin{equation}
    \d^{d-1}p = \frac{2\pi^{\frac{d-1}{2}}}{\Gamma\left(\frac{d-1}{2} \right)}p^{d-2}\d p,
\end{equation}
and the Mellin representation of $a^{-s}$,
\begin{equation}
    a^{-s} = \frac{1}{\Gamma(s)}\int_{0}^\infty \d t \,\, t^{s-1} e^{-ta},
\end{equation}
to rewrite Eq. (\ref{eq: zetad}) as

\begin{eqnarray}
    \zeta_D(s) &=& \frac{2A_{d-1}\pi^{\frac{d-1}{2}}}{(2\pi)^{d-1}\Gamma\left(\frac{d-1}{2}\right)\Gamma(s)} \left(\frac{L^2}{\pi}\right)^s \nonumber \\
    &\times& \int_{0}^\infty \d t \,\, t^{s-1} \sum_{n=1}^{\infty} e^{-tn^2\pi} \nonumber \\
    &\times& \int_0^{\infty} \d p \,\,p^{d-2} \exp\left[\frac{-tL^2}{\pi}\left(p^2 + m_0^2\right)\right].
\end{eqnarray}

The integration over the continuum modes can be readily performed. 
Additionally, we set $m_0^2=0$, because that is the case where the Casimir force 
appears (infinite correlation length), and rename $\zeta_D(s) \rightarrow \zeta_G(s)$,
where $G$~stands for Goldstone. Performing the integral, one obtains for~$\zeta_G(s)$:
\begin{equation}\label{eq:zetapsi}
    \zeta_G(s) = C_d(L,s)\int_{0}^{\infty}\d t\,\,t^{s-\frac{1}{2}(d+1)}\psi(t),
\end{equation}
where we have defined the following quantities
\begin{equation}\label{eq:cs}
    C_d(L,s) \equiv \frac{A_{d-1}}{(2L)^{d-1}\Gamma(s)}\left(\frac{L^2}{\pi}\right)^s,
\end{equation}
\begin{equation}
    \psi(t) \equiv \sum_{n=1}^{\infty}e^{-tn^2\pi}.
\end{equation}
As one can see, the contribution of $\psi(t)$ is rapidly decreasing as $t \to \infty$. 
However, depending on the values of $s$ and $d$, there are singularities at $t\to 0$ 
that need to be taken care of. As discussed in Ref.~\cite{Hawking:1976ja},
the singularity can be removed assuming the system confined to a large, but finite
box, which entails an infrared cutoff in the $p$-integrals above. Instead of introducing
an explicit infrared cutoff, we extract the finite part of the integral by using 
the following relations of $\psi(t)$ and the weight~$1/2$ modular form 
$\Theta(t)$ \cite{riemann1859ueber}:
\begin{equation}
    \psi(t) = \frac{1}{2}\left(\Theta(t) -1\right),
\end{equation}
where
\begin{equation}\label{eq:theta}
    \Theta(t) \equiv \sum_{n\in \mathbb{Z}}e^{-tn^2\pi} \quad \mathrm{and} 
    \quad \Theta(1/t) = \sqrt{t}\,\Theta(t).
\end{equation}
Combining the relation between $\psi(t)$ and $\Theta(t)$ together with the 
modular property of $\Theta(t)$ we can write
\begin{equation}\label{eq:psianal}
\psi(1/t) = t^{1/2} \psi(t) + \frac{1}{2} t^{1/2} - \frac{1}{2}.
\end{equation}
Now we can carry out the analytic continuation of Eq.~(\ref{eq:zetapsi}) with the change of 
variables $t \to 1/t$ and using Eq.~(\ref{eq:psianal}), which leads to
\begin{equation}\label{eq:zeta2}
     \zeta_G(s) =\frac{C_d(L,s)}{2}\left[2I^G_{1,d}(s) + I^G_{2,d}(s) -  I^G_{3,d}(s)\right],
\end{equation}
with $I^G_{1,d}, \dots$ being the integrals:
\begin{align}
    I^G_{1,d}(s) &= \int_{0}^{\infty}\d t\,\,t^{\frac{d}{2} - s - 1}\psi(t), \\
    I^G_{2,d}(s) &= \int_{0}^{\infty}\d t\,\,t^{\frac{d}{2} - s - 1}, \quad \mathrm{and},\\
    I^G_{3,d}(s) &= \int_{0}^{\infty}\d t\,\,t^{\frac{d}{2} - s - \frac{3}{2}}.
\end{align}
%
The integral $I_{1,d}(s)$ 
is convergent for any values of $s$~and~$d$, whereas $I_{2,d}(s)$ diverges for Re$(2s)< d$ and $I^{(3)}_{d}(s)$ diverges for  Re$(2s)< d-1$. 
As can be checked in the Eq. (\ref{eq:cs}), we have that $C_{d}(L,s) \to 0$ as $s \to 0$, 
which implies
\begin{eqnarray}\label{eq:dzeta}
    \left.\frac{\d \zeta_G(s)}{\d s}\right|_{s=0} &=& \frac{1}{2}\left.\frac{\d C_{d}(L,s)}{\d s}\right|_{s=0} \nonumber \\
    &\times&\left[2I^G_{1,d}(0) + I^G_{2,d}(0)  - I^G_{3,d}(0)\right]. \nonumber \\
\end{eqnarray}

The integral $I^G_{1,d}(0)$ is finite, positive definite, does not depend 
on the distance of the plates $L$; it depends only on the dimension $d$, 
and can be performed analytically. 
On the other hand, the divergent integrals $I^G_{2,d}(0)$ and $I^G_{3,d}(0)$ 
do not depend on the distance between the plates and can be dropped considering 
that we have a large box, which implies a large, but finite, wavelength, as argued 
in Ref.~\cite{Hawking:1976ja} and mentioned above. Divergences
would not appear 
if $m_0 \neq 0$. After some simplifications one can obtain that
\begin{align}\label{eq:zetagol}
     \left.\frac{\d \zeta_G(s)}{\d s}\right|_{s=0}&=\frac{A_{d-1}}{(2L)^{d-1}}I^G_{1,d}(0) =\frac{A_{d-1}}{(2L)^{d-1}}\frac{1}{2\pi}\sum_{n=1}^{\infty}\frac{1}{n^{d}}  \nonumber \\ 
     &= \frac{A_{d-1}}{(2L)^{d-1}} \frac{\zeta(d)}{2\pi}.
\end{align}
Using that $F=E-TS$ and the fact that $T=0$ in our case, one concludes
that
\begin{equation}\label{eq:CasE}
    Z = e^{-F} = e^{-E} \Rightarrow E = -\frac{1}{2}  \left.\frac{\d \zeta_G(s)}{\d s}\right|_{s=0}.
\end{equation}
Now we can define the energy density and find that
\begin{equation}\label{eq:enerden}
    \frac{E}{A_{d-1}} \equiv \epsilon_d(L) = -\frac{1}{2(2L)^{d-1}}\frac{\zeta(d)}{2\pi},
\end{equation}
\noindent
which has, evidently, the correct sign and power law with~$L$.
%


 For $d=3$, Eq. (\ref{eq:enerden}) results
\begin{equation}\label{eq:kardar}
   \epsilon_3(L) = -\frac{\zeta(3)}{16\pi L^2},
\end{equation}
which is the ``universal" amplitude of the Goldstone modes \cite{li1991fluctuation}. 
The reason for the quotation marks will become clear at the end of this work. The Casimir 
force per unit of area (Casimir pressure) can be calculated as the negative of the derivative 
with respect to $L$ of Eq. (\ref{eq:enerden}). 

In the next section, we briefly review the technique that will
be used to take into account the disorder, the distributional zeta-function 
method.

\section{Distributional zeta-Function Method}\label{sec:distzeta}

This section aims to review the distributional zeta-function 
method~\cite{Svaiter:2016lha, Svaiter:2016-2}, 
the method we used to obtain the disorder-averaged free energy for 
a system described by statistical field theory or Euclidean quantum 
field theory.
To exemplify the method, we use it to derive the Casimir force for 
a general configuration of the field multiplets, without using a saddle point approximation.

The partition function of the model for one disorder realization in the
presence of an external  source $j(x)$ is given by: 
\begin{equation}
\hspace{-0.2cm}Z(j,h)\!=\!\int\![\d \phi]\, \exp\!\left[ -S(\phi,h) 
\!+\! \int \! \d^{d}x\, j(x)\phi(x)\right]\!,
\label{eq:disorderedgeneratingfunctional}
\end{equation}
%
where the action functional in the presence of additive (linearly coupled) disorder is  
\begin{equation}
S(\phi,h)=S(\phi) + \int \d^{d}x \, h(x)\phi(x).
\label{eq:spe1}
\end{equation}
Here, $S(\phi)$ is the pure system action, and  $h(x)$ is a quenched random field.

In a general situation, one can model a disordered medium by a real random field $h(x)$ in $\mathbb{R}^{d}$ with $\mathbb{E}[h(x)]=0$ and covariance $\mathbb{E}[h(x)h(y)]$, where $\mathbb{E}[\cdots]$ specifies
the mean over the ensemble of realizations of the disorder. Some works have studied the case of a disorder modeled by a complex random field; see Refs.~\cite{sham1976effect, Heymans:2022vqw} and Sec. \ref{sec:main}. As in the pure system case, one can define the system's free energy for one disorder realization $W(j,h)=\ln Z(j,h)$, the generating functional of connected correlation functions for one disorder realization. From $W(j,h)$, one 
can obtain the quenched free energy by performing the average over the ensemble 
of all disorder realizations:
\begin{equation}
\mathbb{E}\bigl[W(j,h)\bigr]=\int\,[\d h]P(h)\ln Z(j,h),
\label{eq:disorderedfreeenergy}
\end{equation} 
where $[dh] = \prod_{x \in \mathbb{R}^d} dh(x)$ is a formal functional measure 
and, $[dh]P(h)$ is the probability distribution of the disorder field. 

For a general disorder probability distribution, the distributional 
zeta-function, $\Phi(s)$, is defined as:
\begin{equation}
\Phi(s)=\int [\d h]P(h)\frac{1}{Z(j,h)^{s}}.
\label{pro1}
\vspace{.2cm}
\end{equation}
For $s\in \mathbb{C}$, this function is defined in the region where the above integral 
converges. One defines the complex exponential $n^{-s}=\exp(-s \log n)$ for $\log n\in\mathbb{R}$.
As proved in Refs.~\cite{Svaiter:2016lha,Svaiter:2016-2}, $\Phi(s)$ is defined for $\Re(s) 
\geq 0$. Therefore, the integral is defined in the half-complex plane, and an analytic continuation 
is unnecessary. We have that 
\begin{equation}
\mathbb{E}\bigl[W(j,h)\bigr] = - \frac{\d \Phi(s)}{\d s}\Bigg|_{s=0^{+}}, \,\,\,\,\,\,\,\,\,\, \Re(s) \geq 0.  
\end{equation}

Using the Euler's integral representation for the gamma function, we get
\begin{equation}
    \Phi(s) = \frac{1}{\Gamma(s)}\int[\d h]P(h)\int_{0}^{\infty}dt\,t^{s-1} e^{-Z(j,h)t}.
\end{equation}
The next step consists in expanding the exponential in the integral in a power series. The series expansion has a uniform convergence for each $h$ in the domain $t \in [0,a]$, where $a$ is a dimensionless arbitrary constant. We then split the integral into two pieces, one that is uniformly 
convergent in the interval $t \in [0,a]$ for $a$~finite, and one 
that becomes small for $a\rightarrow \infty$. The contribution from 
the first piece then becomes a sum over all integer moments of the 
partition function, $\mathbb{E}[Z^k(j,h)]=\mathbb{E}\,[(Z(j,h))^{\,k}]$, while the second vanishes 
exponentially for $a$~large. Explicitly, the average free energy can be represented by the following series 
of the moments of the partition function:
\begin{eqnarray}\label{eq:diszeta}
\mathbb{E}\left[W(j,h)\right]&=&\sum_{k=1}^{\infty} \frac{(-1)^{k+1}a^{k}}{k k!}\,\mathbb{E}\,[Z^k(j,h)]\nonumber  \\
&-& \, \ln(a)-\gamma+R(a,j),
\end{eqnarray}
\noindent 
where $\gamma$ is the Euler-Mascheroni constant, and $R(a,j)$ given by
\begin{equation}
R(a,j)=-\int [\d h]P(h)\int_{a}^{\infty}\,\dfrac{\d t}{t}\, e^{ -Z(j,h)t} .
\end{equation}
For large $a$, $|R(a,j)|$ is small; therefore the dominant contribution to
the average free energy is given by the moments of the partition function of
the model.

For concreteness, we assume a Gaussian form for the probability 
distribution of the disorder field $[dh]\,P(h)$: 
\begin{equation}
P(h) = p_{0}\, \exp\left[ - \frac{1}{2\rho^{2} } \int \! \d^{d}x \; h^2(x)\right],
\label{dis2}
\end{equation}
where $\rho$ is a positive parameter and $p_{0}$ is a normalization constant. In this case, we 
have a delta correlated disorder:
\begin{equation}
\mathbb{E}[{h(x) h(y)}] = \rho^{2}\,\delta^{d}(x-y).
\label{delta-correl}
\end{equation}
After integrating the disorder, one obtains that each moment of the partition function  
$\mathbb{E}\,[Z^{\,k}(j,h)]$  can be written~as:
\begin{equation}\label{eq:kmoment}
\mathbb{E}\,[Z^{\,k}(j,h)]=\int\,\prod_{i=1}^{k} \, [\d \phi_{i}^{k}]\, e^{-S_{\textrm{eff}}(\phi_{i}^{k},j_{i}^{k})},
\end{equation}
where $S_{\textrm{eff}}\left(\phi_{i}^{k},j_{i}^{k}\right)$ is obtained integrating over the disorder field, a standard procedure in the literature~\cite{livro3,livro4}. In the above equation the superscript $k$ in $\phi_i^{k}$ identifies the term of the series expansion given by Eq. (\ref{eq:diszeta}), the subscript $i$ is the component of the $k^{\mathrm{th}}$ multiplet, and $\prod_{i=1}^{k}[d\phi_{i}^{k}]$ represents a product of formal functional measures. Also, from now on, we set $j_i^k(x)=0\,\, \forall \,\, i$ and suppress its appearance as argument of the quantities of interest.

To proceed, we use a Ginzburg-Landau model with $\lambda \phi^4$ interaction. After performing the disorder average, one obtains the effective action:
\begin{eqnarray}\label{eq:effectivehamiltonian}
S_{\text{eff}}(\phi_{i}^{k}) &=& \int \d^{d}x\, \sum_{i=1}^{k}  
\left[\frac{1}{2} \phi_{i}^{k}(x)
\left(-\Delta\,+m_{0}^{2}\right)\phi_{i}^{k}(x)\right. \nonumber \\
&-& \frac{\rho^2}{2}\sum_{i,j=1}^k \phi_{i}^{k}(x)\phi_{j}^{k}(x) 
+ \left.\frac{\lambda}{4}\!
\sum_{i=1}^{k} \left(\phi_{i}^{k}(x)\right)^4\right]\!.
\end{eqnarray}
The $\phi^4$ term is necessary to stabilize a ground state of the system since
the disorder average introduces a negative contribution, quadratic in the
fields.
For simplicity, we assume \emph{in this section} the ansatz 
$\phi_{i}^{k}(x)=\phi_{j}^{k}(x)$ for the function space; in which case the
effective action becomes: 
\begin{eqnarray}\label{Seff-II}
S_{\text{eff}}(\phi_{i}^{k}) &=& \int \d^{d}x\, \sum_{i=1}^{k}  
\left[\frac{1}{2} \phi_{i}^{k}(x)
\left(-\Delta\,+m_{0}^{2} - k\rho^2\right)\phi_{i}^{k}(x)\right. \nonumber \\
&+& \left.\frac{\lambda}{4}\!\sum_{i=1}^{k} 
\left(\phi_{i}^{k}(x)\right)^4\right]\!.
\end{eqnarray}
Such a simplified ansatz  has been studied in several works using this method \cite{Diaz:2016mto, Diaz:2017grg, Diaz:2017ilf, Soares:2019fed, Rodriguez-Camargo:2021ryf, Rodriguez-Camargo:2022wyz, Heymans:2022sdr} and leads to consistent results.
Very recently~\cite{Heymans:2023tgi}, we have shown that one can avoid such a simplification and work with the full set of arbitrary field configurations $\{\phi_{i}^{k}(x)\}$. 
For now, to explain the zeta-distributional method to compute the Casimir energy, we proceed with the simplified ansatz.

One sees in Eq.~(\ref{Seff-II}) that there exists a combination of $m_0^2,\,k$ and, $\rho$ 
for which $m_0^2 - k \rho^2 < 0$, signaling the
spontaneous breaking of the discrete symmetry 
$\phi^k_i \rightarrow - \phi^k_j$. As usual, one can move from 
the ``false" vacuum to the ``true" vacuum by an appropriate shift of 
the fields and identify the mass in the Gaussian contribution to
the action:
\begin{equation}
    m^2_\rho \equiv 2(k\rho^2 - m_0^2) > 0 .
\end{equation}
To discuss the Casimir energy, it is enough to consider the 
Gaussian contribution. This is so because, 
as shown by several studies within quantum field theory scenarios~\cite{Wieczorek:1986cw, 
Robaschik:1986vj, Kong:1997iq, melnikov2001radiative}, radiative corrections are always subleading compared to the free-field contribution. 
Since the critical Casimir effect studied here 
is formally identical to the quantum scalar case, the scenario is the same. 
Therefore, we drop the non-Gaussian 
terms in the action. Now, compactifing one dimension and 
assuming Dirichlet boundary conditions, one can recast 
the mean over the $k$-th moment, Eq.~(\ref{eq:kmoment}), as
\begin{equation}
    \mathbb{E}\,[Z^{\,k}(h)]= \left[\mathrm{det}(-\Delta + m_\rho^2)_{\Omega_L}\right]^{-\frac{k}{2}}.
\end{equation}

From now on, we consider the situation  $m_\rho^2 > 0$. Using the spectral
zeta-function regularization, Sec. \ref{sec:speczeta}, we can write the functional determinant as:
\begin{equation}
    \mathbb{E}\,[Z^{\,k}(h)] = \exp\left[\frac{k}{2}\left.\frac{\d }{\d s}\zeta_\rho(s)\right|_{s=0}\right].
\end{equation}
The $\zeta_\rho(s)$ can be constructed as
\begin{equation}
    \zeta_\rho(s) = \frac{A_{d-1}}{(2\pi)^{d-1}}\int \d^{d-1}p \sum_{n=1}^{\infty} \left[p^2 + m_\rho^2 + \left(\frac{\pi n}{L}\right)^2 \right]^{-s}.
\end{equation}
Following the same steps as those between 
Eqs. (\ref{eq: zetad}) and (\ref{eq:dzeta}), but for a 
nonzero mass, we obtain:
\begin{eqnarray}\label{eq:dzeta2}
    \left.\frac{\d \zeta_ \rho(s)}{\d s}\right|_{s=0} &=& \frac{1}{2}\left.\frac{\d C_{d}(L,s)}{\d s}\right|_{s=0} \nonumber \\
    &\times&\left[2I_{1,d}^\rho(0) + I_{2,d}^\rho(0) - I_{3,d}^\rho(0)\right], 
\end{eqnarray}
with
\begin{eqnarray}
I_{1,d}^\rho(s)  &=& \int_{0}^{\infty}\d t\,\,t^{\frac{d}{2} - s - 1}
e^{\frac{-L^2m_\rho^2}{\pi t}}\psi(t), \label{eq:rho1} \\
I_{2,d}^\rho(s)  &=& \int_{0}^{\infty}\d t\,\,t^{\frac{d}{2} - s - 1}
e^{\frac{-L^2m_\rho^2}{\pi t}},\\
I_{3,d}^\rho(s)  &=& \int_{0}^{\infty}\d t\,\,t^{\frac{d}{2} - s - \frac{3}{2}}
e^{\frac{-tL^2m_\rho^2}{\pi}} \label{eq:rho4}.
\end{eqnarray}

Since now we have a nonzero mass, all integrals 
are convergent. Some care must be taken to define the energy of the system. First of all, we 
recall that at zero temperature, the quenched free 
energy can be written as
\begin{eqnarray}
    F_q(L) &=& E_q(L) = -\mathbb{E}\left[W(j,h)\right] \nonumber \\ 
    &=& \sum_{k=1}^{\infty} \frac{(-1)^{k}a^{k}}{k k!}\,\mathbb{E}\,[(Z(j,h))^{\,k}].
\end{eqnarray}

Using the previous results and exponentiating the $a^k$, 
we obtain the Casimir energy in the presence of quenched disorder, 
for now on we call such a quantity as \textit{quenched Casimir} energy,
\begin{equation}
    E_q(L) = \sum_{k=k_c}^{\infty} \frac{(-1)^{k}}{k k!} \exp\left[k\ln a + \frac{k}{2}\left.\frac{\d }{\d s}\zeta_\rho(s)\right|_{s=0}\right],
\end{equation}
with $k_c$ defined as
\begin{equation}
    k_c \equiv \left \lfloor \frac{m_0^2}{\rho^2} \right \rfloor,
\end{equation}
where $\lfloor x \rfloor$ is the greatest integer less than or equal to $x$.

Analysing the behavior of the integrals 
Eq. (\ref{eq:rho1})-(\ref{eq:rho4}), it is immediate to see that for each $k>k_c$ the exponential 
damping makes their contributions sub-leading. 
So the main contribution in the expression for the 
Casimir energy will be
\begin{equation}\label{eq:mainener}
    E_q(L) = \frac{(-1)^{k_c}}{k_c k_c!} \exp\left[k_c\ln a + \frac{k_c}{2}\left.\frac{\d }{\d s}\zeta_\rho(s)\right|_{s=0}\right].
\end{equation}
Clearly, from the 
last equation, we can see the connection between $a$ and the 
thermodynamic limit: since $\zeta_\rho(s)$ is an extensive 
quantity, $a$ must be chosen to maximize the exponential. 
Therefore, the Casimir force is given by:
\begin{equation}\label{eq:fcas}
    f_d(L) \equiv -\frac{\partial E_q(L)}{\partial L} = \frac{(-1)^{k_c + 1}}{2 k_c!} \frac{\partial}{\partial L}\left.\frac{\d }{\d s}\zeta_\rho(s)\right|_{s=0}.
\end{equation}
With the results obtained up to now, 
we have that
\begin{eqnarray}\label{eq:force}
    &\hspace{-0.2cm}&f_d(L) = \frac{A_{d-1}}{2^{d+1}}\frac{(-1)^{k_c + 1}}{ k_c!}\nonumber \\
    &\hspace{-0.2cm}& \times\left\{ -\frac{1}{L^d}\left[2I_{1,d}^\rho(0) + I_{2,d}^\rho(0)  - I_{3,d}^\rho(0)\right]  \right. \nonumber \\
    &\hspace{-0.2cm}& +  \frac{L^{1-d}}{d-1}\left. \frac{\partial}{\partial L}\left[2I_{1,d}^\rho(0) + I_{2,d}^\rho(0) - I_{3,d}^\rho(0)\right]\right\}. \nonumber \\
\end{eqnarray}
The derivative of $I_{i,d}^\rho$ deserves a closer look. All of those integrals have an exponential which depends on $L^2$ and, 
thanks to the exponential and the $\psi(t)$ term, their derivatives 
with respect to $L/2$ do not change 
their convergence properties.
In a power series expansion in $L/2$,
the contribution of the second term of Eq. (\ref{eq:force}) 
has a global contribution proportional 
to $-L^{2-d}$, which ensures that such a contribution 
is the leading one in powers of $L/2$. Now, defining the quenched 
Casimir pressure as the quenched Casimir force per unit 
area ($d-1$ volume), we can write
\begin{equation}\label{eq:presure}
    p_d(L) = \frac{(-1)^{k_c}}{2^{d+1} k_c!L^{d}}\left[\frac{L^2}{d-1}B_d(0) + D_d(0)\right],
\end{equation}
where $B_{d}(0)$ and $D_{d}(0)$ are defined by

\begin{eqnarray}   
B_{d}(0) &\equiv& -\frac{1}{L}\frac{\partial}{\partial L} 
    \Bigl[2I_{1,d}^\rho(0) + I_{2,d}^\rho(0) - I_{3,d}^\rho(0)\Bigr], \\
    D_{d}(0) &\equiv&  2I_{1,d}^\rho(0) + I_{2,d}^\rho(0) - I_{3,d}^\rho(0),
\end{eqnarray}
are positive constants. Clearly, for $m_\rho^2=0$ the $B_{d}(0)$ 
vanishes and the well know behavior is recovered. The most interesting 
feature of Eqs. (\ref{eq:fcas}) and 
Eq. (\ref{eq:presure}) is the fact that the factor of $(-1)^{k_c}$ 
can change the force or pressure from repulsive to attractive
depending on the values of $m_0^2$ and $\rho^2$.
In the next section we further explore such a feature. 
 Alongside considering the breaking of a 
continuous symmetry breaking, which creates soft modes 
in the system, we also do not make any ansatz over the function space.

\section{Interplay between soft and critical modes}
\label{sec:main}

In order to verify and go beyond the results of Ref. \cite{Rodriguez-Camargo:2022wyz}, 
we now consider a system with a continuous symmetry $U(1)\cong O(2)$.
Another difference will be in the function space that we obtain after taking
the average of the logarithm of the partition function.
To start, let us consider the action
\begin{eqnarray}
    S(\phi,\phi^*) &=& \frac{1}{2}\int \d^{d} x \left[\phi^*(x)\left( - \Delta + m_0^2\right) \phi(x) +\lambda V(\phi, \phi^*) \right.\nonumber \\
    &+& \left. h^*(x)\phi(x) + h(x)\phi^*(x)\right];
\end{eqnarray}
as before, $m_0^2$ is the bare mass, $\lambda$ is a strictly positive constant 
and $V(\phi, \phi^*)$ is a polynomial in the field
variables. Here we would like to point out that in the case of interacting field theories confined in compact domains, is necessary to introduce surface counterterms \cite{Symanzik:1981wd, Diehl:1981zz, Fosco:1999rs, Caicedo:2002ft, AparicioAlcalde:2005wxe}
The main difference here is that $h(x)$ is now a 
complex random field \cite{yaglom1957some,sham1976effect, Heymans:2022vqw}, 
with a probability distribution $P(h,h^*)$.
Again, to simplify the problem, we consider a Gaussian distribution 
\begin{equation}
    P(h,h^*) \equiv p_0e^{-\frac{1}{\rho^2}\int \d^dx\left|h(x)\right|^2}.
\end{equation}
The $k$-th moment in the series, Eq.~(\ref{eq:kmoment}),
with $j(x) = 0$, generalizes to:
\begin{equation}\label{eq:moment}
    \mathbb{E}\left[ Z^k(h) \right] = \int \prod_{i,j=1}^k [\d \phi^k_i][\d \phi^{k*}_j] \; e^{ -  S_{eff}(\phi^k_i, \phi_j^{k*})},
\end{equation}
with
\begin{equation} 
S_{eff}(\phi^k_i, \phi^{k*}_j) = \sum_{i,j} \left[S_0 (\phi^k_i, \phi^{k*}_j) 
+ \lambda S_I(\phi^{k}_i, \phi^{k*}_j) \right].
\end{equation}
Here, $S_0(\phi^k_i, \phi^{k*}_j)$ is the quadratic action:
\begin{align}\label{eq:eff0}
    S_0(\phi^k_i, \phi^{k*}_j) &= \frac{1}{2} \int \d^d x \,\,
    \phi^{k*}_i(x) \left(G^0_{ij} - \rho^2\right)\phi^{k}_j(x),
\end{align}
in which, for later convenience, we defined
\begin{equation}
    G^0_{ij} \equiv \left(-\Delta + m_0^2\right)\delta_{ij},
    \label{G0ij}
\end{equation}
and $S_I(\phi_i, \phi^*_j)$ is the interaction action 
corresponding to $V(\phi, \phi^*)$. 
The propagator 
corresponding to $S_0(\phi_i, \phi^*_j)$ is not diagonal in the 
$(i,j)$-space. Such a nagging feature 
has been previously dealt with in different 
ways in the literature. For example, one can work with 
a non-diagonal propagator, as in some of the minimal supersymmetric 
standard model extensions~\cite{Lewandowski:2017omt,Lewandowski:2018bnn}, 
or one can use a Hubbard-Stratonovich identity as in the 
Bose-Hubbard model~\cite{sachdevbook}. 
Still another way is to use the ansatz $\phi^k_i = \phi^k_j$,
as discussed in the last section. Although such an ansatz leads to 
consistent results, it is an unnecessary simplification 
as one can use the spectral theorem of linear algebra to formally 
diagonalize the propagator~\cite{Heymans:2023tgi}. This 
diagonalization is a new development in the distributional 
zeta-function method, introduced in Ref.~ \cite{Heymans:2023tgi} 
in a different context.

The diagonalization proceeds as follows. We define
the matrix of the $k\times k$ propagator as
\begin{equation}
    G \equiv \left[\begin{array}{cccc}
        G^0_{11} - \rho^2 & -\rho^2 & \cdots & -\rho^2  \\
   -\rho^2 & G^0_{22} - \rho^2 & \cdots & -\rho^2  \\
   \vdots & \cdots & \ddots & \vdots \\
   -\rho^2 & -\rho^2 & \cdots & G^0_{kk} - \rho^2  \\
  \end{array}\right]_{k\times k} ,
  \label{G-matrix}
\end{equation}
where $G^0_{ij}$ was defined in Eq.~(\ref{G0ij}).
Since $G$ is 
a symmetric matrix,  
it can be diagonalized by an orthogonal 
matrix~$S$ whose columns are the eigenvectors of $G$:
\begin{equation}\label{eq:D}
    D = \langle S, GS\rangle,
\end{equation}
where $\langle,\rangle$ denotes 
the natural inner product in $(i,j)$-space, and $D$ is 
the (diagonal) matrix of eigenvalues of $G$. Using the vector $\Phi(x)$ as the 
vector which has components $\phi_i(x)$, we can 
rewrite the sum of the quadratic
actions as  
\begin{align}
    \sum_{i,j=1}^k S_0 (\phi_i, \phi^*_j) & = \frac{1}{2}\int \d^dx \,\, \langle\Phi(x), G \Phi^*(x)\rangle \nn \\
    &= \frac{1}{2}\int \d^dx \,\, \langle\tilde{\Phi}(x), D \tilde{\Phi}^*(x)\rangle,
\label{S0-phitilde}
\end{align}
where $\tilde{\Phi}(x) = S \Phi(x)$ and 
\begin{equation}
    D = \left[
\begin{array}{cccc}
  G^0_{11} -k\rho^2 & 0& \cdots & 0 \\
   0 & G^0_{22} & \cdots & 0 \\
   \vdots & \cdots & \ddots & \vdots \\
   0 & \cdots & & G^0_{kk}
\end{array}
\right]_{k\times k}.
\end{equation}
The matrix $S$ can be calculated exactly; due to the degeneracy 
of the spectrum, there is many matrices that diagonalizes $G$. The components of $\tilde{\Phi}(x)$ will
be given by a linear combination  of the $\phi_i(x)$ 
determined~by~$S$. 
Let $\varphi_i(x)$ denote the components of 
$\tilde{\Phi}(x)$ by $\varphi_i(x)$.  Using the component 
notation, one can 
write  the diagonal form of the quadratic action 
in Eq.~(\ref{S0-phitilde}) as
\begin{align}
\label{eq:actions}
      &\sum_{i,j=1}^k S_0 (\phi_i, \phi^*_j) =  \frac{1}{2}\int \d^dx \,\, 
      \varphi^*(x)(-\Delta + m_0^2 - k\rho^2) \varphi(x) \nonumber \\
      &\hspace{1.65cm} +   \frac{1}{2} \sum_{a=1}^{k-1} \int \d^dx \,\, 
      \varphi^*_a(x)(-\Delta + m_0^2) \varphi_a(x) ,
\end{align}
where, to simplify the notation henceforth, we defined 
$\varphi_1(x) \equiv \varphi(x)$ and also changed
the dummy index 
in the second line. Since $S$ is 
an orthogonal matrix, one has that
\begin{equation}\label{eq:mes}
    \prod_{i,j=1}^k [\d \phi_i][\d \phi^*_j] = [\d \varphi][\d \varphi^*]\prod_{a,b=1}^{k-1} [\d \varphi_a][\d \varphi^*_b]. 
\end{equation}
Therefore, using Eqs. (\ref{eq:actions}) and (\ref{eq:mes}) into  Eq. (\ref{eq:moment}), we obtain:
\begin{eqnarray}\label{eq:partf}
     \hspace{-0.5cm}\mathbb{E}\left[ Z^k(h) \right] &=& \int[\d \varphi][\d \varphi^*]\prod_{a,b=1}^{k-1} [\d \varphi_a][\d \varphi^*_b]\nonumber \\
     &\times& e^{-S_\rho(\varphi, \varphi^*) - \sum_{a}S_O(\varphi_a, \varphi_a^*) 
     - \lambda S_I(\varphi_a, \varphi_a^*)},
\end{eqnarray}
where $S_\rho(\varphi, \varphi^*)$
is the action carrying the information on the strength~$\rho$ of the disorder,
\begin{equation} 
S_\rho(\varphi, \varphi^*) = \frac{1}{2}\int \d^dx \, 
    \varphi^*(x)(-\Delta + m_0^2 - k\rho^2) \varphi(x), 
\label{eq:Srho}
\end{equation}
and $S_O(\varphi_a, \varphi_a^*)$ is a $O(k-1)$-symmetric action, 
independent of the disorder, given by:
\begin{equation}
S_O(\varphi_a, \varphi_a^*) =  \frac{1}{2} \int \d^dx \,
   \varphi^*_a(x)(-\Delta + m_0^2) \varphi_a(x) .
\label{eq:SO}
\end{equation}
The action $S_I(\varphi_a, \varphi_a^*)$ will not be
needed in our study of the Casimir effect, but its presence with
a $\lambda > 0$ is required to guarantee the action boundness. Its
explicit expression is readily obtained by replacing $\Phi$ in the
original action by $\tilde\Phi = S\Phi$.

We proceed recalling 
that each moment of the partition function 
contributes to the total quenched free energy, 
Eq. (\ref{eq:diszeta}). To obtain the Casimir energy 
we compactify one of the 
dimensions,  $\mathbb{R}^{d} \to \mathbb{R}^{d-1}\times [0,L]$, and impose some boundary 
conditions. 
As can be seen in 
Eq. (\ref{eq:Srho}), there is a combination of $k,m_0^2$ and $\rho$ 
for which the effective mass $m^2_0 - k \rho^2$  
becomes negative, indicating the symmetry breaking 
$U(1) \to \mathbb{Z}_2$, giving rise to a 
Goldstone (soft) mode. 
Of course, the Casimir force is present even for those terms 
in the sum with a positive effective mass, as the condition for its presence
is that the correlation length becomes of the order of the 
system's compactified size~$L$. That is, the total
energy receives contributions from symmetry-preserving and symmetry-breaking
terms. Our 
interest in this work is study the interplay
between the contributions to the energy of the 
symmetry-breaking soft mode and the critical mode, both induced 
by the disorder. Therefore we neglect the symmetry-preserving modes. We assess
this interplay by first performing a shift in the field $\varphi(x)$ to 
expose the symmetry breaking, then neglect all non-Gaussian terms, and
finally, take the large~$L$ limit. 

We perform the  
symmetry-breaking field shift for the situation with
 $m_0^2 - k\rho^2 < 0$ in Eq. (\ref{eq:Srho}). 
In the Cartesian representation of the fields $\varphi(x)$ and 
$\varphi^*(x)$ we have that
\begin{eqnarray}
    \varphi(x) &=& \frac{1}{\sqrt{2}} \left[ \psi_1(x) 
    + i \psi_2(x)\right], \\
     \varphi^*(x) &=& \frac{1}{\sqrt{2}} \left[ \psi_1(x) - i \psi_2(x)\right].
\end{eqnarray}
The minima of the action lie on the  
circle  
\begin{equation}
    \psi_1^2 + \psi_2^2 = \frac{2(k\rho^2-m_0^2)}{\lambda} 
    \equiv v^2.
\end{equation}
Definining the shifted fields 
$\chi = \psi_1 - v$ and $\psi = \psi_2$, 
the Gaussian part of the action becomes
\begin{eqnarray}
    S_{\rho}(\chi,\psi) &=& \frac{1}{2}\int \d^dx \left[\chi(x)(-\Delta + m_\rho^2)\chi(x) \right. \nonumber \\
    &+& \left. \psi(x)(-\Delta)\psi(x)\right],
\end{eqnarray}
where we defined $m_\rho^2 = 2(k\rho^2-m_0^2)$. In the new variables, after dropping  
all non-Gaussian terms, Eq. (\ref{eq:partf}) assumes 
the following enlightening form:
\begin{equation}
    \mathbb{E}\left[ Z^k(h) \right] 
    = Z_{\rho}Z_{G}\left[Z_{O}\right]^{k-1},
    \label{Z-k}
\end{equation}
where
\begin{eqnarray}
     Z_{\rho} &=& \int [\d \chi] \; e^{-\frac{1}{2}\int \d^dx \,\chi(x)(-\Delta + m_\rho^2)\chi(x)}, \label{eq:Zrho}\\
     Z_{G} &=& \int [\d \psi] \; e^{-\frac{1}{2}\int \d^dx \, \psi(x)(-\Delta)\psi(x)}, \label{eq:ZG} \\
     Z_{O} &=& \int [\d \varphi] [\d \varphi^*] \; e^{-\frac{1}{2}\int \d^dx\, \varphi^*(x)(-\Delta + m_0^2)\varphi(x)},\label{eq:Zo}
\end{eqnarray}
where the partion functions are respectively the contributions of the disorder, the Goldstone mode, and a $O(k-1)$ symmetric model.

Now,
we take a slab geometry with one compactified dimension, $\Omega_L =  \mathbb{R}^{d-1}\times [0,L]$, and impose Dirichlet boundary conditions to all fields
\begin{equation}
    A_\alpha(x_1, \cdots, x_{d-1}, 0) = A_\alpha(x_1, \cdots, x_{d-1}, L) = 0,
\end{equation}
with $\alpha = \{\rho, G, O\}$ and $\{A_\rho,A_G,A_O\} = \{\chi, \psi, \varphi\}$ respectively. Using the result in 
Eq. (\ref{eq:fundet}) for each of the 
partition functions in Eqs.~(\ref{eq:Zrho}), (\ref{eq:ZG}), and
(\ref{eq:Zo}), we obtain for the $k$-th moment of the 
partition function, Eq.~(\ref{Z-k}), the following expression:
\begin{eqnarray}
    \mathbb{E}\left[ Z^k(h) \right] &=& \left[\mathrm{det}(-\Delta + m_\rho^2)_{\Omega_L}\right]^{-\frac{1}{2}}\left[\mathrm{det}(-\Delta)_{\Omega_L}\right]^{-\frac{1}{2}}\nonumber \\ &\times&\left[\mathrm{det}(-\Delta + m_0^2)_{\Omega_L}\right]^{-\frac{k-1}{2}}. 
\end{eqnarray}
The last term contributes neither to the critical nor 
to the soft Goldstone modes. As such, it can be dropped 
by redefining the energy.

The relevant contributions to the Casimir energy can be 
regularized using the spectral 
zeta regularization
\begin{equation}
    \mathbb{E}\left[ Z^k(h) \right] = \exp\left\{\left.\frac{1}{2}\frac{\d}{\d s}\left[\zeta_\rho(s) + \zeta_G(s)\right]\right|_{s=0}\right\}.
\end{equation} 
By the same arguments used to obtain Eq. (\ref{eq:mainener}) in 
the previous section, one concludes  
that the main contribution to the total quenched Casimir energy is
given by
\begin{equation}
    \hspace{-0.25cm}E^T_c = \frac{(-1)^{k_c }}{k_ck_c!}
    \exp\left\{k_c\ln a + \left.\frac{1}{2}\frac{\d}{\d s} 
    \left[\zeta_\rho(s) + \zeta_G(s)\right]\right|_{s=0}\right\}.
\end{equation}

We define the following zeta function
\begin{equation}
    \zeta_\alpha (s) = \frac{A_{d-1}}{(2\pi)^{d-1}}\int \d^{d-1}p \sum_{n=1} \left[p^2 + m_\alpha^2 + \left(\frac{\pi n}{L}\right)^2 \right]^{-s},
\end{equation}
with $\alpha = \{\rho, G\}$ and $m_G^2 = 0$. Using the same definitions and arguments
in Sec. \ref{sec:speczeta}, one can rewrite 
$\zeta_\alpha (s)$ as
\begin{equation}
     \zeta_\alpha (s) = C_{d}(L,s)\int_{0}^{\infty} \d t \,\, t^{s-
     \frac{1}{2}(d+1)} e^{\frac{-tL^2}{\pi}m_\alpha^2}\psi(t).
\end{equation}
Following the same steps taken between 
Eqs. (\ref{eq:zetapsi}) and (\ref{eq:zeta2}), it
is straightforward to obtain that
\begin{equation}
    \zeta_{\alpha}(s) = C_{d}(L,s) \left[2I^\alpha_{1,d}(s) + I^\alpha_{2,d}(s) - I^\alpha_{3,d}(s)\right],
\end{equation}
where

\begin{eqnarray}
    I^\alpha_{1,d}(s) &=&\int_{0}^{\infty} \d t \,\, t^{\frac{d}{2}-s-1} e^{\frac{-L^2}{\pi t}m_\alpha^2}\psi(t), \\
    I^\alpha_{2,d}(s) &=& \int_{0}^{\infty} \d t \,\, t^{\frac{d}{2}-s-1} e^{\frac{-L^2}{\pi t}m_\alpha^2}, \\
    I^\alpha_{3,d}(s) &=& \int_{0}^{\infty} \d t \,\, t^{\frac{d}{2}-s-\frac{3}{2}} e^{\frac{-tL^2}{\pi}m_\alpha^2}.
\end{eqnarray}

One obtains the quenched Casimir force analogously 
to Eq. (\ref{eq:force}). Such a force 
receives contributions from 
the spectral zeta functions of soft and critical modes. 
In the case of $\alpha = G$ we have the same situation  
of Sec. \ref{sec:speczeta} for $m_0 = 0$, \textit{i.e.}, 
the contribution of the soft modes to the Casimir force 
is given by Eq.~(\ref{eq:zetagol}).
For $\alpha = \rho$, we have the calculation of Sec. \ref{sec:distzeta} 
and the corresponding contribution is given by Eq. (\ref{eq:dzeta2}). Putting all 
together{\color{red},} we obtain for the total quenched Casimir pressure of the 
system {\color{red}the} following expression:
\begin{eqnarray}
\label{eq:final}
p^T_d(L) &=&  \frac{(-1)^{k_c }}{k_ck_c!2^{d-1}L^d} \left[\frac{L^2}{d-1}B_d(0) 
+ D_d(0) + \frac{\zeta(d)}{2\pi}\right].\nonumber \\
\end{eqnarray}
 Such a result can be plotted as function of $L$ for different dimensions and values of $k_c$. Figures~1 and 2 
 a  display $p^T_d(L)$ for dimensions 2,3, and 4 for 
different values of $k_c$. Note the different scales in the axes of the two figures.
\begin{figure}[ht!]
\includegraphics[scale=0.74]{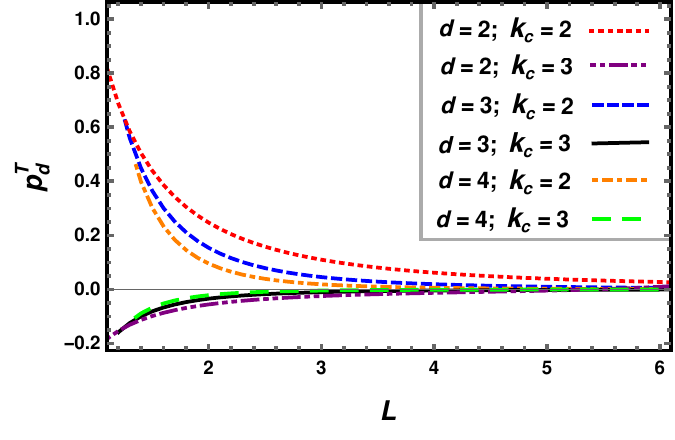}
\label{fig:2}
\caption{Plot of the quenched Casimir pressure, Eq. (\ref{eq:final}), for dimensions $2,3,$ and $4$ and $k_c= 2$, and $3$.}
\end{figure}

\begin{figure}[ht!]
\includegraphics[scale=0.74]{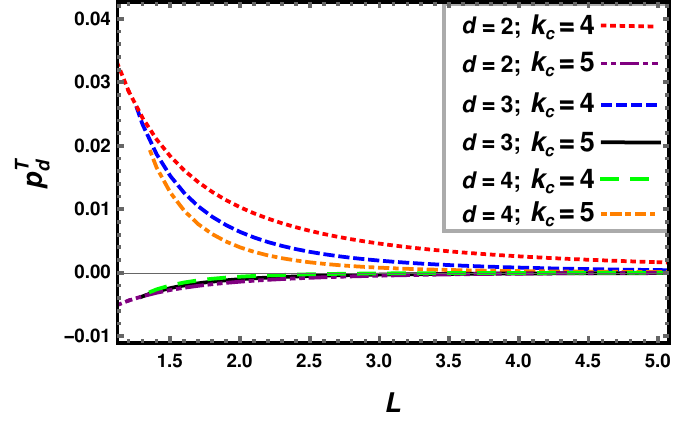}
\label{fig:3}
\caption{Plot of the quenched Casimir pressure, Eq. (\ref{eq:final}), for dimensions 
$2,3,$ and $4$ and $k_c= 4$, and $5$.}
\end{figure}

This result has some interesting features. 
First of all, if we ignore the 
Goldstone mode contributions, the resulting equation 
 differs from Eq. (\ref{eq:presure}) by
a multiplicative factor, $4/k_c$. This factor comes from 
the exact diagonalization of the quadratic actions; when one 
uses the ansatz $\phi_i^k(x) = \phi_j^k(x)$ $\forall \,i,\,j$,  
as used in Sec. \ref{sec:distzeta}, the multiplicative factor 
does not appear. Of course, such a difference is irrelevant to gathering
qualitative understanding. However, the qualitative similarity between the results
holds only when one can neglect the contribution from the partition function 
$Z_O$, Eq.~(\ref{eq:Zo}). This is the case whenever 
the corresponding action 
does not reach criticality,
a situation that can occur due to nonzero temperature or finite-size effects.  
Another feature 
of Eq. (\ref{eq:final}) is that the critical and the soft 
mode effects are noncompetitive, they are of the same sign.
Still another interesting feature is that, when $k_c\rho \gg m_0^2$, 
one can neglect the contribution of $Z_\rho$, Eq.~(\ref{eq:Zrho}), to the 
Casimir energy; in practice, one can set $B_d(0)=D_d(0)=0$ in Eq.~(\ref{eq:final}). 
This is interesting because then only soft modes 
contribute, but with a factor proportional to $(-1)^{k_c}$, which means that 
a change of sign may occur. 
In other words, there is a universal constant due to the soft modes, given by 
$\zeta(3)/16\pi$, with an overall sign that 
can be either negative (as usual) or positive, depending on the value of $k_c$. 

\section{Conclusions}\label{sec:conc}

In this work we analyzed the interplay in the Casimir energy between the 
soft modes from 
the breaking of a continuous symmetry, and the critical modes, due to a 
disorder linearly coupled to a complex scalar field. 
We found that both modes always 
have a cooperative effect making the quenched Casimir pressure stronger. 
More interesting we have seen that, 
in the scenario of strong disordered system, the Goldstone mode 
contribution to the pressure can be either
positive or negative, depending on the ration between the strengh of disorder and mass parameter. This fact can be relevant 
in stability analyses of systems at 
nano scales, where those effects are expected to be larger
than one atmosphere \cite{dantchev2023critical}. 

From a technical point of view,
in this work we made use of a significant improvement on 
the application of the zeta distributional method regarding the 
functional space of fields. The functional space is ansatz-free, 
in the sense that we have not made any special choices on the 
fields in the nondiagonal effective action resulting from the disorder 
averaging. Moreover, we have made use of the spectral theorem of linear algebra to formally
diagonalize the effective action in the full functional space. 
These features seem to be applicable to any Gaussian theory, bosonic or 
fermionic.

Further topics in the critical Casimir effect in disordered systems which 
deserve attention 
include: analyses 
 on how boundary shape and temperature and/or 
finite-size effects may affect the procedure that 
we described here. In addition, it would be interesting to extend the ``ansatz-free" 
approach to interacting field theories, both for additive and multiplicative 
disorder. These subjects are under investigation by the authors.

\section*{Acknowledgments} 
The authors are grateful to S. A. Dias for fruitful discussions, and also to the referees,  whose comments and suggestions lead to a better presentation of our work. 
This work was partially supported by Conselho Nacional de Desenvolvimento Cient\'{\i}fico e Tecnol\'{o}gico
(CNPq), grants nos. 305894/2009-9 (G.K.), 305000/2023-3 (N.F.S.), and 311300/2020-0 (B.F.S.), INCT 
F\'{\i}sica Nuclear e Apli\-ca\-\c{c}\~oes, grant no. 464898/2014-5 (G.K.), Funda\c{c}\~{a}o de Amparo \`{a} 
Pesquisa do Estado de S\~{a}o Paulo (FAPESP), grant no. 2013/01907-0 (G.K.), and Fundação Carlos Chagas 
Filho de Amparo à Pesquisa do Estado do Rio de Janeiro (FAPERJ) grant no. E-26/203.318/2017 (B.F.S.). G.O.H. 
thanks to Fundação Carlos Chagas Filho de Amparo à Pesquisa do Estado do Rio de Janeiro (FAPERJ) due the 
Ph.D. scholarship.  


%

\end{document}